\documentclass[12pt,notitlepage]{article}

\newcommand{\bd}{
\begin{document}}
\newcommand{\ed}{\end{document}}
\newcommand{\beq}{\begin{equation}}
\newcommand{\eeq}{\end{equation}}
\newcommand{\su}{\section}
\newcommand{\ssu}{\subsection}
\newcommand{\sssu}{\subsubsection}
\newcommand{\nid}{\noindent}
\newcommand{\bea}{\begin{eqnarray}}
\newcommand{\eea}{\end{eqnarray}} 
\newcommand{\baR}{\begin{array}}
\newcommand{\eaR}{\end{array}}
\newcommand{\ben}{\begin{enumerate}}
\newcommand{\een}{\end{enumerate}}
\newtheorem{proposition}{Proposition}
\newcommand{\bp}{\begin{proposition}}
\newcommand{\ep}{\end{proposition}}
\newtheorem{conjecture}{Conjecture}
\newcommand{\bconj}{\begin{conjecture}}
\newcommand{\econj}{\end{conjecture}}

\newcommand\bbbz{{\sf Z\!\!Z}}                                            
\newcommand\bbbn{{\sf I\!N}}
\newcommand\bbbr{{\sf I\!R}}

\newcommand\hX{\hat{\bX}}
\newcommand\hY{\hat{\bY}}
\newcommand\hF{\hat{\bF}}
\newcommand\hm{\hat{\mu}}
\newcommand\be{{\bf e}}
\newcommand\bE{{\bf E}}
\newcommand\bF{{\bf F}}
\newcommand\bJ{{\bar J}}
\newcommand\bX{{\bf X}}
\newcommand\bY{{\bf Y}}
\newcommand\bU{{\bf U}}
\newcommand\bZ{{\bf Z}}
\newcommand\cA{{\cal A}}
\newcommand\cF{{\cal F}}
\newcommand\cG{{\cal G}}
\newcommand\cC{{\cal C}}
\newcommand\cM{{\cal M}}
\newcommand\cP{{\cal P}}
\newcommand\cS{{\cal S}}
\newcommand\cT{{\cal T}}
\newcommand\cW{{\cal W}}

\newcommand\hM{\hat{\cm}}
\newcommand\hcp{\hat{\cP}}
\newcommand\ml{\mu_L}
\newcommand\hml{\hat{\mu}_L}
\newcommand\cm{\cM}
\newcommand\dL{^\partial \Lambda}
\newcommand{\deq}{\stackrel {\rm def}{=}}

\newcommand\ta{{\tilde a}}
\newcommand\tom{\tilde{\omega}}

\bd

\pagestyle{plain}

\title {Some fractal aspects of Self-Organized Criticality.}

\author {B. Cessac\\Institut Non Lin\'eaire de Nice\\ 1361 Route des Lucioles, 06560 Valbonne, France.}
\date{}
\maketitle

\def\abstractname{Abstract}
\begin{abstract}
The concept of Self-Organized Criticality (SOC)  was proposed in an attempt to explain
the widespread  appearance of power-law in nature. It describes a mechanism in which a  system
reaches spontaneously a state where the characteristic events (avalanches) are 
distributed according to a power law.
We present a dynamical systems approach to Self-Organized Criticality where the
dynamics is described either
in terms of Iterated Function Systems, or as a piecewise hyperbolic
dynamical system of skew-product type. Some results linking
the structure of the attractor and some characteristic properties
of avalanches are discussed.
\end{abstract}

\su{Introduction}

The notion of Self-Organized Criticality (SOC)  
has become a new paradigm  to explain the widespread appearance of power-law  in a multitude
of examples like the distribution of the size of earthquakes, 1/f-noise, amplitudes
 of solar flares, species extinction .... to name only a  few cases \cite{Bak,Jensen}.
In this paradigm,
the dynamics occurs as chain reactions or \textit{avalanches}.
A stationary regime is reached where
the distribution of avalanches follows a power
law, namely there is scale invariance reminiscent of thermodynamic systems at
the critical point.  A local perturbation
can induce effects at any scale and there are
long-range spatial and time correlations.
 In other words, in this paradigm
the system  reaches {\it spontaneously} a
critical state without  any fine tuning of some control parameter.
Several models have been proposed like the sandpile model
\cite{Bak1,Bak2}, the abelian sandpile \cite{Dhar} or the continuous energy model 
\cite{Zhang}. The results available are mainly numerical and
only a few rigorous results are known. Moreover, the scientific community is still
lacking a general formalism to handle these models properly.\\

It is natural 
 to try to access the macroscopic behavior of large sized systems
from the microscopic dynamical evolution. In this spirit,
we have developed a dynamical system description for a certain class of SOC models
(like Zhang model \cite{Zhang}), for which the whole SOC dynamics can either be described
in terms of Iterated Function Systems, or as a piecewise hyperbolic 
dynamical system of skew-product type.  Several deep
 results from the theory of hyperbolic dynamical systems can then be used, having interesting
 implications on the SOC dynamics. The general setting and some of these results are discussed in this paper.

\su{Definitions.}\label{Def}

In this paper we consider the Zhang model which is defined as follows. 
Let $\Lambda$ be a d-dimensional box in $\bbbz^d$, taken as a
cube of edge length $L$ for simplicity and let $\dL$
be the boundary of $\Lambda$, namely the set of points
in $\bbbz^d \setminus\Lambda$ at distance $1$ from $\Lambda$.
 Call $N=\#\Lambda=L^d$,
where $\#$ denotes the cardinality  of a set. 
Each site $i \in \Lambda \cup \dL$
is characterized by its
"energy" $X_i$, which is a non-negative real and finite number. 
An \textit{energy configuration} is a vector $\bX  = \lbrace
X_i \rbrace_{_{i \in \Lambda}}$. The sites of $\dL$ have always
zero energy : as discussed below this mimics energy dissipation at the boundaries.
Let $E_c$ be a real, positive  number, called the 
\textit{critical energy}, and $\cM = [0,E_c[^N$.
An energy configuration $\bX$ is called \textit{stable} when $\bX \in \cM$
and \textit{unstable} otherwise. 
In an unstable configuration the sites
 $i$, such that $X_i \geq E_c$, are called \textit{active} or unstable.
The dynamics  on $\bX$ 
depends whether $\bX$ is stable
or unstable.

If $\bX$ is stable,
one chooses a site $i\in \Lambda$ at random with probability 
$\frac{1}{N}$, and add to it the energy
$\delta  = 1$ (\textit{excitation}).
If $\bX$ is unstable,
each active site  loses a part of its energy,
redistributed in equal parts to its $2d$ neighbors 
in the following way (\textit{relaxation}).
Fix $\epsilon \in [0,1[$ and set $\alpha=\frac{(1-\epsilon)}{2d}$. 
When $i$ is active 
 it gives the energy $\alpha X_i$ to its $2d$ neighbors
and keeps the energy $\epsilon X_i$.
Therefore, the energy is locally conserved during relaxation. The  relaxation dynamics is synchronous and 
it is useful to express it in terms of the  map:
\beq \label{F}
\bF(\bX)=\bX +\alpha\Delta\left[\bZ(\bX)\ast\bX \right].   
\eeq
\nid where  $\bZ(\bX)$ is a $N$ dimensional vector,
such that $Z_i(\bX)=0$ if $X_i<E_c$
and $Z_i(\bX)=1$ if $X_i \geq E_c$.
The $\ast$ denotes the product component by component:
 if $\bX,\bY$ are $N$ dimensional vectors,
$\bX \ast \bY$ is the $N$ dimensional vector of components
$X_i Y_i$.  $\Delta$ is the discrete Laplacian on $\Lambda$
 with zero boundaries conditions. Due to the presence of thresholds,
 $\bF$ corresponds to energy propagation by \textit{singular} diffusion.\\

Zhang model dynamics consists therefore in resting periods where energy
is injected and stored locally (excitation), followed by periods of
activity called \textit{avalanches}: when the energy of a given site exceeds the
threshold, this energy is partially released and distributed to the neighbors.
This leads to a chain reaction that can propagate on very long scales. 
However, the energy reaching the boundaries is lost  since we imposed the boundaries to have always a zero energy. 
This mimics a dissipation mechanism which implies 
that all avalanches stop within a {\it finite} number of
iterations. Note that with these updating rule, each avalanche starts from {\it only one}
active site.

The structure of an avalanche
is encoded by the sequence of active sites:
$$\cC = \left\{C_t\right\}_{1 \leq t \leq \tau_\cC},$$
\nid where: 
$$
C_t = \left\{j \in \Lambda | X_j \geq E_c \ \mbox{in the t-th step
of avalanche}\right\},
$$
\nid and where $\tau_\cC$, called the \textit{avalanche
duration}, is the smallest positive integer such that $C_{\tau_\cC+1}=\emptyset$. Each avalanche can be labeled by a double index
$\left( i
,j \right)$. The first index refers to the site
where the energy is dropped and the second index labels the different
avalanches starting at $i$ (including the ``empty'' avalanche where the
excitation of $i$ does not render it active).  Moreover, to each avalanche $(i,j)$ corresponds  a convex domain
$\cm_{(i,j)}$ in $\cm$. For each $i$, the domains $\cm_{(i,j)}$ form a partition of $\cm$ \cite{BCK3}.

The total energy of a stable configuration in a finite lattice being finite,
(it is bounded by $L^d E_c$), the total number of different avalanches 
is finite for finite $L$,
(but diverges as $L \to \infty$).
 The \textit{size} $s$ of an avalanche
is the total number of active sites, 
$s(\cC)=\#\cup_t C_t$. The \textit{area} $a$ is the number
of distinct active sites in $\cC$. We will generically denote by $n$ 
an avalanche observable (size, duration, area) and $n(i,j)$ will be
the value that $n$ takes in the avalanche $(i,j)$. \\

It is numerically observed that, after a sufficiently long time, the dynamics
reaches an out of equilibrium stationary state\footnote{A precise definition 
will be given below.} where the energy injected
per unit time is equal, on average, to the energy dissipated at the boundaries.
In this regime, the avalanche observables are distributed according to a truncated power law:
\beq \label{PL}
P_L(n)=\frac{C_L}{n^{\tau}}f_L(n), \ 1 \leq n\leq \xi_L
\eeq
Here and in the sequel, the subscript $L$ will refer to the size of $\Lambda$.
$C_L$ is a normalization constant depending on $L$, and $\xi_L$ is 
the maximal value that the observable can take in a box of size $L$ (note that
this quantity depends also on $E_c,\epsilon$). As discussed above $\xi_L$ is finite
for finite $L$, but it diverges (typically like $L^\beta$) when $L \to \infty$.
 $\tau$ is called the \textit{critical exponent}
of the avalanche observable.  $f_L(n)$ is a finite size cut off, accounting for boundaries
effect, and
such that  $\lim_{L \to \infty} f_L(n) =1, \ \forall n \leq \xi_L$.
Consequently, as $L \to \infty$, $P_L(n)$ converges to a power law distribution $P^\ast(n)=\frac{K}{n^{\tau}}$.

\su{Avalanche maps.}

Each  avalanche $(i,j)$  maps  a stable
configuration $\bX \in \cm_{(i,j)}$ to the next stable configuration 
obtained when injecting the energy $\delta$ at site $i$. Consequently, 
to each avalanche $(i,j)$ we associate a map $T_{(i,j)} : \cm_{(i,j)} \to \cm$
such that :
\beq
T_{(i,j)}(\bX) = \bF^{\tau(i,j)}(\bX+\delta \be_i), \quad \bX \in \cm_{(i,j)},
\eeq
\nid where $\be_i$ is the $i$-th canonical basis vector of $\bbbr^N$.
Note that $T_{(i,j)}$ is a simple translation in the case where
the excitation of $i$ does not render it active.\\

 Let $\cA$ be the set of all
avalanches symbols $(i,j)$. The transition $(i_1,j_1) \to (i_2,j_2)$ is \textit{legal}
iff:
\beq
T_{(i_1,j_1)}\left[\cm_{(i_1,j_1)}\right] \cap \cm_{(i_2,j_2)} \neq \emptyset
\eeq  
This means that some energy configuration $\bX \in \cm_{(i_1,j_1)}$
can undergo the avalanche $(i_2,j_2)$ after the avalanche $(i_1,j_1)$.
Note that since the excited sites are selected randomly and independently,
there is no constraint in the choice of the symbol $i_2$.
In other words, there are at least $N$ legal transitions for each symbol
$(i,j)$.  The \textit{avalanche transition graph} $\cG_\cA$ is the graph of
legal transitions.
It is tempting to use the avalanche coding for symbolic coding
of the dynamics. However, the coding is in general not one to one.\\

The maps $T_{(i,j)}$ have some remarkable properties \cite{BCK3}:

\ben
\item[(i)] Each $T_{(i,j)}$ is \textit{affine}:
\beq
T_{(i,j)}(\bX)=L_{(i,j)}(\bX)+C_{(i,j)}
\eeq
\nid where $L_{(i,j)}$ is an $N \times N$ matrix and $C_{(i,j)}$ a constant.

\item[(ii)] Each $T_{(i,j)}$ is a \textit{quasi-contraction}:
\beq
\rho(L_{(i,j)})\leq 1
\eeq
\nid where $\rho$ is the spectral radius. Moreover, the number of eigenvalues
of modulus strictly lower than one is equal to the avalanche area $a(i,j)$

\item[(iii)] The determinant of $L_{(i,j)}$ (local volume contraction) is directly related 
to the avalanche size $s(i,j)$ by:
\beq \label{Contract}
det L_{(i,j)} = \epsilon^{s(i,j)}
\eeq
\item[(iv)] There exists $t_0 \equiv t_0(L,E_c,\epsilon) < \infty$ and $\eta < 1$, such that, 
for any legal path $\gamma_{t_0}$ of length $t_0$ on $\cG_\cA$ one has:
\beq\label{ContractProd}
\rho(L_{\left[\gamma_{t_0}\right]}) < \eta < 1
\eeq 
\nid where $L_{\left[\gamma_{t_0}\right]}$ is the composition of the matrices
$L_{(i,j)}$ along $\gamma_{t_0}$. Namely, each composed map $T_{\left[\gamma_{t_0}\right]}$
is a contraction.
\een

It follows from these properties that Zhang model can be described as a 
probabilistic graph directed iterated function systems.

\su{Dynamical system.}

Zhang model  can also be described as an hyperbolic dynamical system of skew product type,
with singularities. The phase space is the set  $\hM =
\Sigma^+_\Lambda \times \cM$,  where $\Sigma^+_\Lambda$ is
the set of right infinite excitation sequences 
$\ta = \left\{ a_1,\dots, a_k, \dots \ | a_k \in \Lambda
\right\}$ such that $a_t$ is the $t$-th excited site. A point in $\hM$ is denoted by $\hX =(\ta,\bX)$.
We call $\pi^+$ (resp. $\pi^-$) the projection on $\Sigma^+_\Lambda$ 
(resp. on $\cm$) such that $\pi^+(\hX)=\ta$ and $\pi^-(\hX)=\bX$.

The dynamic evolution is given by the map $\cT : \hM \to \hM$
such that:
\beq \label{SD}
 \cT(\hat{\bX}) \deq \left( \sigma \ta, T_{a_1}(\bX) \right)
\eeq 
\nid where $\sigma$ is the left shift on $\Sigma^+_\Lambda$.
$T_{a_1}$ is the mapping from $\cm  \to \cm$ whose restriction to $\cm_{a_1,k}$ is $T_{(a_1,k)}$.\\

The shift $\sigma$ is conjugated to $Nx \ mod \ 1$ on $[0,1]$. Therefore, the dynamical system (\ref{SD}) has a positive Lyapunov
exponent $\lambda_L(0)=log(N)$. Due to the property (\ref{ContractProd}) it has also $N$ strictly negative Lyapunov exponents
$\lambda_L(i),\quad   i =1 \dots N$ corresponding to the projection of the dynamics (\ref{SD}) on $\cm$.
The maps $T_{a_1}$ are discontinuous at the boundaries of the domains $\cm_{a_1,k}$. The union
of these boundaries, $\cS= \cup_{a_1=1^N}\cup_{k=1}^{n(a_1)} \partial \cm_{a_1,k}$ is called the 
\textit{singularity set}. This is the set of energy configurations such that at least one site has an energy exactly equal
to $E_c$. Therefore,
 the slightest change in its energy can change dramatically the further evolution.
The singularity set plays therefore an important role (see \cite{BCKM} for a discussion).\\

As discussed above Zhang model is expected to reach a stationary state when time tends to infinity.
In the context of the dynamical system (\ref{SD}) it is natural to define this state
as the Sinai-Ruelle-Bowen (SRB) measure. This is the weak limit:
\beq \label{SRBmix}
\hml=\lim_{t\to \infty}  \cT^{*t}(\mu)
\eeq
\nid where $\mu$ is the Lebesgue measure on $\hM$. Note that the existence of the
limit (\ref{SRBmix}) in Zhang model can be rigorously established only in some restricted case \cite{BCK3}.
However this is a natural assumption from a physicist point of view, since it means that the stationary regime
does not depend on the way the system is prepared.
It is supported by numerical simulations and also by some rigorous results established in \cite{BCK1,BCK3,BCKM}.

\su{Symbolic coding.}

As discussed above, the avalanche coding is not suitable for symbolic coding.
However, it is possible to construct a finer partition allowing a one to one
correspondence between an infinite sequence of symbols and a point in $\cm$.
 More precisely, it can be shown, in some  cases, that the dynamical system 
 (\ref{SD}) has a finite Markov partition. We conjecture\footnote{This conjecture is based 
on the following (numerical) observation \cite{BCKM}.
The $\hml$ measure of the $\eta$-neighborhood of the  singularity set $\cS$ decreases like 
$\eta^\alpha$ where $\alpha \leq 1$. From this property and from the Borel-Cantelli lemma it follows
that Lebesgue almost every point in $\hM$ has a local stable
manifold of positive diameter. Each point has also an unstable manifold (corresponding to
the direction of the shift in the extended phase space).
This implies the existence of a finite Markov partition.}
that this is true for generic values of $E_c$. \\

Call $\hcp= \left\{\hcp_\omega\right\}_{\omega \in \Omega}$ this partition,
where $\Omega$ is the finite set of symbols parameterizing the Markov
partition elements..
By construction each element $\hcp_\omega$ has a product structure
$\pi^+(\hcp_\omega) \times \pi^-(\hcp_\omega)$. The projection $\pi^+(\hcp_\omega)$ of $\hcp_\omega$
on $\Sigma^+_\Lambda$ is a cylinder set consisting of all sequences
starting with a definite $a_1 \in \Lambda$, while the projection $\pi^-(\hcp_\omega) \subset \cm$.
For simplicity, we will use the notation $\cP_\omega = \pi^-(\hcp_\omega)$

To each element $\omega$  
corresponds a sub-domain $\cP_\omega \subset \cm_{ij}$
and henceforth a \textit{unique  avalanche} $(i,j)$. Equivalently, to
each $\omega$ one can associate a double index $\left( i\left( \omega \right)
,j\left( \omega \right) \right)$ where the first index 
refers to the site
where the avalanche starts and the second to
the corresponding avalanche. 
Note however that, in general, \textit{several symbols}
correspond  to the \textit{same avalanche} since
 a domain $\cm_{ij}$ is usually  composed by several partition-elements $\cP_\omega$.
However, in order to simplify the notation we will 
 identify $\omega$ and the avalanche
$\left(i(\omega),j(\omega)\right)$ whenever this causes
no confusion. To summarize, we have now a symboling coding useful both for the dynamics \textit{and} for labeling
the avalanches.\\ 

Moreover, the evolution of a probability distribution on $\Omega$ 
can be encoded in $\textit{Markov transition}$ graph with a transition matrix $\cW$ such that:
\beq
\cW_{\omega \omega'} = \hml\left[ \cT^{-1}(\hcp_{\omega'}) | \hcp_\omega \right]
\eeq 
It can be shown in some cases that $\cW$ is irreducible \cite{BCK3}.
As discussed above (see the comments about the weak limit (\ref{SRBmix}))
it is natural to expect that this is a generic situation. Consequently, the corresponding Markov
process admits
a unique invariant measure $m_L$ such that $m_L= m_L\cW$. The induced measure on the 
space of legal right infinite sequence on $\Omega$,
$\Sigma_\Omega^+$, is conjugated to the SRB measure (\ref{SRBmix}) \cite{BCKM}.

\su{Main results.}

We have now two possible pictures for Zhang model. This is a probabilistic graph directed
iterated function system with invariant probability $m_L$, where each map
$T_\omega$ has a volume contraction $\epsilon^{s(\omega)}$.  Note however
that the maps are not conformal. Under the assumption of irreducibility of the graph,
there is a unique attractor. As discussed below it is expected to be fractal for sufficiently
small $E_c$ values.
Zhang is also an hyperbolic dynamical system
of skew product type, with singularities. Both aspects are fruitful. 
In this section, we discuss the main results obtained from the formalism developed above.

\ssu{Average contraction rate and avalanche size.}

It follows from eq. (\ref{Contract}) and the ergodic theorem that :
\beq \label{contav}
\sum_{i=1}^N \lambda_L(i)=\log(\epsilon)\left<s\right>_{L}.
\eeq
\nid where  $\left<s\right>_{L}$ is the average avalanche size.
There is therefore a strong connexion between the microscopic volume contraction and 
the average avalanche size.

More generally, it follows from eq. (\ref{Contract}) and (\ref{PL})
that the local volume contraction of the maps $T_\omega$
on a typical trajectory is distributed according to a truncated power law
of type (\ref{PL}). This suggests therefore a strong connexion between
the fractal properties of the stationary state and the avalanche size distribution
(\cite{BCKM,C}).

\ssu{Hausdorff dimension versus $E_c$.}

The following proposition can easily be proved.

\bp \label{musing}
$\hml$ is singular for all $E_c$ sufficiently small.
\ep

Indeed, the $L_1$ norm of all map $T_\omega$ is bounded from above by one minus the energy flux
dissipated at the boundaries \cite{BCK4}. This flux $\in [0,1]$  and tends to $1$ when $E_c \to 0$.
Consequently, for $E_c <<1$ one can make the $L_1$ norm of all map $T_\omega$ arbitrary
small. 
Since the expansion is constant it follows that $det(D\cT) < 1$ hence $\hml$ is singular (but, as a SRB measure,
it is absolutely continuous along the unstable foliation).  

When $E_c$ increases, the average contraction rate decreases since the smaller is the average avalanche size
$<s>_L$. Therefore, for fixed $N$, there exists an $E_c^*(N)$ which is the unique $E_c$ value
such that:
\beq \label{Ec*}
\log(\epsilon) <s>_L + \log(N) = 0
\eeq
For $E_c < E_c^*(N)$ the contraction dominates the expansion, while it is the opposite for
$E_c > E_c^*(N)$. Clearly, the invariant set structure is  different
 in these two cases. On the one hand, for small $E_c$ values, the images of the domains $\hcp_\omega$ are thin bands which
are stretched slower than they contract. Therefore,  the invariant set has a Cantor structure with large gaps. On the other hand, when $E_c > E_c^*(N)$,
 the successive images of the domains $\cP_{\omega}$  fill more
and more  the phase space. Consequently, the Hausdorff dimension of the attractor 
is expected to increase for increasing $E_c$, $E_c < E_c^*(N)$ and is likely to be constant when
$E_c > E_c^*(N)$ .

We conjecture the following:

\bconj
The Hausdorff dimension of $\mu_L$ is piecewise continuous and monotonously increasing on the domains
of continuity.
\econj

This is supported by the following argument. On open intervals $I_i$ of $E_c$ the structure of the mappings 
$T_{(i,k)}$ does not change but the domains of continuity $\cM_{(i,k)}$ do. Furthermore for $E_c$ decreasing the 
probabilities for avalanches with higher contraction increases. This  should force the Hausdorff
dimension to decrease monotonously with decreasing $E_c$ on each $I_i$.

\ssu{Multifractal spectrum and energy transport.}

The maps $T_\omega$ are not conformal.  In this case, the multifractal
spectrum is obtained from a (sub-additive) thermodynamic formalism \cite{Falconer3}.
The corresponding potential is a function of the singular values 
$\alpha_i(\tom,k), \ i=1 \dots
N, k=1 \dots \infty$, of the product matrix 
$L_{\omega_k}.\dots L_{\omega_1}$ along a legal sequence $\tom \in \Sigma_\Omega^+$.

It is interesting to remark that these singular values are directly related
to the eigenvalues of the tangent map $D\bF_\bX$ of the relaxation map (\ref{F}). On the other hand,
one can shown that $D\bF_\bX$ is the evolution operator of a Markov process that governs
the propagation of  energy in Zhang model \cite{BCK3,CM2}. The 
eigenmodes of $D\bF_\bX$  for the singular diffusion (\ref{F}) are the analog of Fourier modes  
for normal diffusion. The eigenvalues of $D\bF_\bX$ define a hierarchy of characteristic times.
where the effective energy transport has distinct characteristics: singular on short time
scales, anomalous at intermediate scales, and normal on long time scales
\cite{BCK3,CM2}. Consequently, there is a close connexion between the energy transport
and the multifractal spectrum of $\hml$. Some aspects of this connexion, involving
Lyapunov exponents, have been discussed in \cite{BCK4}. Further developments are under investigation
\cite{CM2}.

\ssu{Ledrappier-Young formula and critical exponent of avalanche sizes.}

The Ledrappier-Young formula \cite{LedYou} relates the positive Lyapunov exponents,
the corresponding partial Hausdorff dimensions and the Kolmogorov-Sinai entropy.
Applied to the backward dynamics of Zhang model it writes:

\beq \label{LY}
-\sum_{i=1}^N \lambda_L(i) \sigma_L(i) =   \log(N) - \log(J_N)
\eeq

\nid where  the $\sigma_L(i)$ are the partial Hausdorff dimensions
in the direction $i$. Let  $J_N(\hat{\bX})$ be the number of preimages of $\hat{\bX}$ then
 $J _N= \int J_N(\hat{\bX}) d\mu(\hat{\bX})$ is the averaged number of preimages. $\log(N) - \log(J_N)$
is therefore the backward entropy. In the case where the dynamics is invertible $\log(J_N)=0$
and the backward entropy is equal to the forward entropy. 
 When the system is not invertible, one
can  make it  invertible
by coding the backward iteration tree in the same way as we did with the excitation sequences,
hence introducing an additional variable on which the forward dynamics contracts. 

Except in some very restricted cases \cite{BCK3} $\log(N) - \log(J_N)$ is proportional to $\log(N)$.
Since the partial Hausdorff dimensions obey $\sigma_L(i) \leq 1$ the equation (\ref{LY}) implies
$-\sum_i \lambda_L(i) \geq \alpha \log(N)$. Then, from eq. (\ref{contav}), we obtain an upper  bound
on the avalanche size:
\beq
<s>_L \geq \frac{\alpha}{|\log(\epsilon)|} \log(N)
\eeq
\nid with strict equality iff all the partial dimensions are equal to $1$. 
When $\hml$ is singular $<s>_L$   diverges therefore faster than logarithmically with $N$. 
This  implies that the critical exponent of $P_L(s)$ is such that $\tau < 2$. We get therefore a bound on the critical
exponent of avalanche size from fractal considerations.
\su{Conclusion.}

We have shown that certain classes of models of SOC like the Zhang model fit
naturally into a well known class of dynamical systems. Especially for  the
question of asymptotic energy distribution, observables distribution,
ergodicity,  this seems to be a proper point of view.  Furthermore it
 exhibits close relationship between the probability  of the size of
avalanches and the fractality of the attractor. Further developments are under investigations.

\ed